# Development and Testing of an Engineering Model for an Asteroid Hopping Robot

Greg Wilburn[a]*, Himangshu Kalita[b], Jekan Thangavelautham[c]

[a] *Department of Aerospace and Mechanical Engineering, University of Arizona, 1130 N. Mountain Ave., Tucson, Arizona, USA 85719*, gregwilburn@email.arizona.edu
[b] *Department of Aerospace and Mechanical Engineering, University of Arizona, 1130 N. Mountain Ave., Tucson, Arizona, USA 85719*, hkalita@email.arizona.edu
[c] *Department of Aerospace and Mechanical Engineering, University of Arizona, 1130 N. Mountain Ave., Tucson, Arizona, USA 85719*, jekan@email.arizona.edu
* Corresponding Author

**Abstract**

The science and origins of asteroids is deemed high priority in the Planetary Science Decadal Survey. Two of the main questions from the Decadal Survey pertain to what the "initial stages, conditions, and processes of solar system formation and the nature of the interstellar matter" that was present in the protoplanetary disk, as well as determining the "primordial sources for organic matter." Major scientific goals for the study of planetesimals are to decipher geological processes in SSSBs not determinable from investigation via in situ experimentation, and to understand how planetesimals contribute to the formation of planets. Ground based observations are not sufficient to examine SSSBs, as they are only able to measure what is on the surface of the body; however, in situ analysis allows for further, close up investigation as to the surface characteristics and the inner composure of the body. The Asteroid Mobile Imager and Geologic Observer (AMIGO) is a 1U stowed autonomous robot that can perform surface hopping on an asteroid with an inflatable structure. It contains science instruments to provide stereo context imaging, micro-imaging, seismic sensing, and electric field measurements. Multiple hopping robots are deployed as a team to eliminate single-point failure and add robustness to data collection. An on-board attitude control system consists of a thruster chip of discretized micro-nozzles that provides hopping thrust and a reaction wheel for controlling the third axis. For the continued development of the robot, an engineering model is developed to test various components and algorithms. Three enabling technologies for the mission are tested. One of the primary components is the inflatable structure that enables context imaging, communication with a mother spacecraft, and solar collection. The other two components tests are for a small reaction wheel system and the MEMS thruster assembly. The inflatable, once properly deployed, is filled with helium to provide a buoyant force simulating micro-gravity conditions and the attitude control system is tested. One algorithm to be tested is organized motion planning to efficiently explore the surface of a simulated asteroid. To enable this path planning, the stereo camera must provide context imaging and the system autonomously determines a point of interest to hop to.

## 1. Introduction

In-situ analysis on an asteroid's surface is required to extend the limited science data obtainable through Earth-based observations. Telescopes can reveal information on the bulk characteristics of small solar system bodies like spectral type, bulk density and composition, their dynamic nature, and approximate orbital history [1]. However, there remain basic unknowns of asteroids including cohesion of the outer surface regolith, electrostatic forces, thermal effects, and geologic structure [2].

In conjunction with a larger scale "mother" spacecraft, small surface landers that can sample asteroid properties at multiple surface locations can provide robust science data. These low-scale robots aim to achieve significant science contribution by simplifying their goals and diversifying. To sample multiple locations, the robots need some form of mobility through either roving, internally actuated hops, mechanical hops, or propulsive hops.

*1.1 Other Robots*

On the surface of an asteroid, local gravity is low enough that traditional rovers are not able to achieve mobility. A wheeled rover relies on traction from friction, a force proportional to the normal force from the robot's weight. The lower the frictional force, the smaller the traction must be which means the rover's wheels must revolve incredibly slowly; otherwise, the wheels would slip with no translation of the rover. Besides, the surface roughness and simple act of spinning a wheel would likely send the robot into a tumble with the potential to reach escape velocity [3]. Thus, wheeled rovers are a non-option.

Internally actuated devices typically rely on spinning up and braking a reaction wheel. A benefit to this system is the actuators are shielded from the surface regolith, extending their lifetime and limiting the probability of failure. The dynamics of the surface regolith must be well understood for accurate prediction of hopping dynamics, as the force transferred to the robot is dependent on robot-regolith interaction. Hedgehog is one such development by NASA JPL and Stanford, with three flywheels and external spikes to tumble for short distances and hop for more distant targets [4,5]. Another is the Gyrover that contains spinning flywheels attached to a two-link manipulator [6]. A recent successful example of this concept is JAXA's MINERVA-II 1A and 1B landers, as they landed and hopped around the surface of Ryugu and transmitted images (Fig. 1).

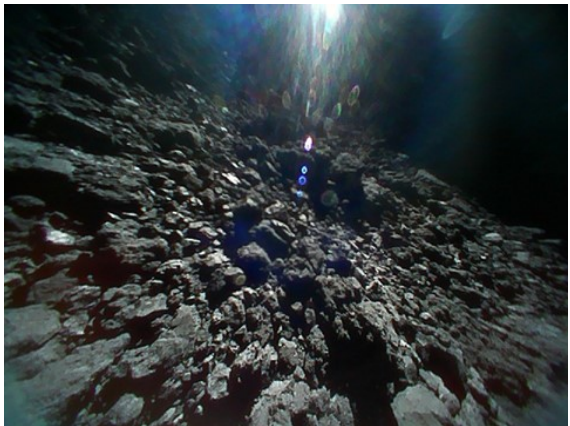

Fig. 1. Surface of Ryugu from MINERVA-II 1B (JAXA, University of Tokyo)

One type of mechanical hopping is by the use of a spring mechanism, a direct reactive force pushing the robot from the surface. The Canadian Space Agency developed the Micro-hopper for traversing Martian terrain, though with a limitation of only one hop per day due to the time to reform the shape memory alloy [7]. Another technique for hopping developed by Plante and Dubowsky at MIT utilize Polymer Actuator Membranes (PAM) to load a spring. The system is only 18 grams and can enable hopping of Microbots with a mass of 100 grams up to 1 m [8].

Another example is SPIKE, a 75 kg spacecraft-hopper that embeds science instruments into regolith via a boom connected to the robot in free fall [9]. Vibrating the boom causes cohesion with regolith to be broken and the spacecraft is free to hop to another location. Again, mechanical hoppers have a reliance on surface characteristics, which are not well constrained and vary asteroid to asteroid.

Thrusters allow for mobility independent of surface characteristics, though exhaust may cause interference with the electrically charged, organic regolith and kick up dust in the process. Another example is the Sphere-X, a spherical robot that hops using chemical propulsion and is intended for exploring in higher gravity of 1.0 m/s2 and higher [10,11,12,13]. This system, however, relies on reaction wheels to provide attitude control, as the thruster is used only for launching the robot. A thruster with multiple nozzles is required for pointing authority for smaller robots with less volume and mass for angular momentum transfer devices.

## 2. AMIGO Mission Concept

The Asteroid Mobile Imager and Geologic Observer (AMIGO) (Figs. 2-3) is a conceptual 1U CubeSat to explore the surface after deployment from a mother spacecraft in orbit around the target body. The 10×10×10 cm chassis has the capacity to hold avionics, micro-propulsion system, a deployable inflatable structure, a stereo camera, a micro camera, and a seismic sensor. The inflatable structure deploys to up to 1 meter in diameter with the stereo camera mounted on top.

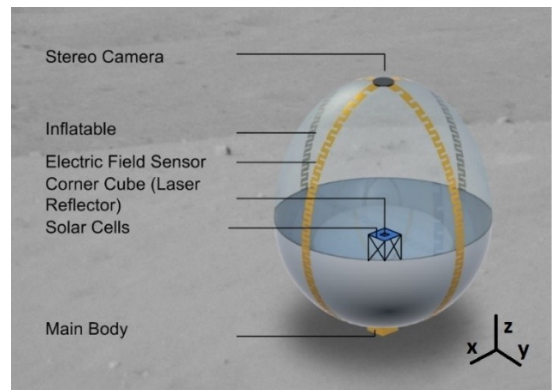

Fig. 2. AMIGO System Overview

The stereo camera provides context imaging and surface mapping/ path planning capability. The context of where the robot is allows for the robot to determine points of interest on the asteroid surface. It also provides imaging for the large-scale surface structure, akin to Fig. 1. The depth mapping capability of stereo imaging allows for path planning for the robot to hop to multiple locations for robust sampling.

The inflatable structure is a critical multi-functional device. By mounting the stereo camera on top, less dust will cling to the lens for clearer and longer-term imaging while providing a larger range of viewing. The inflatable also serves as an antenna to communicate with the mother spacecraft for science data and positioning [14, 24]. The inflatable is also easy to see from the overhead orbiter for positioning updates and reducing the possibility of losing the robot in the potentially deep regolith. Flexible photovoltaic cells could be interlaced into the thin-film structure, allowing for energy collection to extend mission lifetime with lighter batteries.

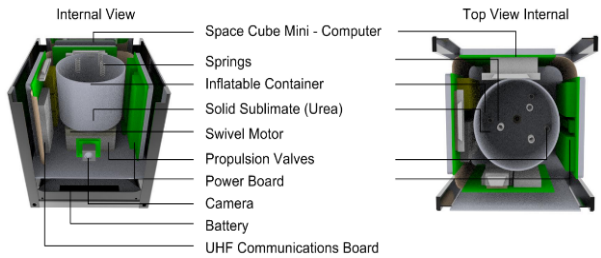

Fig. 3. AMIGO Internals

*2.1 Concept of Operations*

Each AMIGO is deployed from a mother spacecraft (Fig. 4). During descent, the robot inflates from its stowed 1U state. The bottom-heavy design facilitates upright landing, though the inflatable is designed to withstand the slow ~15 cm/s impacts. Upon landing, initial context is determined for where the robot is on the surface. This is done by both on board imaging and tracking from the mother. The inflatable portion provides a tracking target, as smaller robots may not be large enough to be tracked. From there, the science mission is conducted. For the AMIGO lander, there are five science goals:

1. Determine local surface hardness and compliance
2. Acquire seismic data constraining the geologic competence of the asteroid
3. Acquire micro-imaging of fine geologic structure from diverse locations
4. Detect images of thermal fatigue of surface rocks
5. Measure electric fields and properties of surface regolith

Each of these science goals seeks to fill a current knowledge gap in the characteristics of asteroids. For example, the proposed NASA Asteroid Redirect Mission was to retrieve a boulder from the surface of a near Earth asteroid and return the sample for further analysis [15]. Currently, the dynamics of how to extract a boulder from the surface of an asteroid is an open problem. The issue is as fundamental as Newton's Third Law; if one aims to pull a three-ton boulder from the asteroid surface, the spacecraft must exert three tons on the asteroid. Will the asteroid and boulder have enough cohesive strength to not completely fall apart? Seismic sensors and close-up geologic sensors will provide this information. The top-mounted camera provides context to determine local areas of interest and potential locations to traverse to.

The characterization of surface regolith of asteroids is vital to the success of future lander missions and the further understanding of the composition of asteroids. For instance, it is theorized that planetesimals often impacted with each other and either obliterated into fine dust and small clumps or aggregated together. In either case, fine grains are created. For intact planetesimals, this dust accreted to the surface and became the surface regolith. However, that regolith may not have the same compositions as the asteroid itself due to being a combination of multiple meteoric impact events. In situ analysis will aid in the understanding of the surface of asteroids in this regard. A large reason for the concept of AMIGO is to add to the current base of knowledge for the surface characteristics of asteroids for use in future lander missions. The familiarity with asteroid surfaces gained by lower cost missions will lay the foundation for, say, a Discovery class mission to be more successful due to limiting the unknowns in the geology dynamics of asteroids.

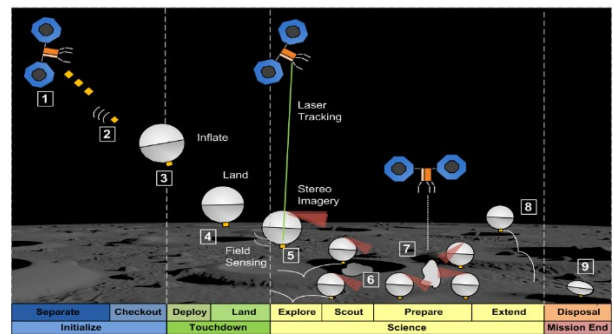

Fig. 4. AMIGO CONOPS

*2.2 Control Actuation*

Motion of the robot is obtained by two types of actuators: an array of micro-thrusters and a reaction wheel. The thruster array is a MEMS chip of micro-nozzles based on sublimate cold gas propulsion (Fig. 5) [16]. The purpose of the propulsion system is to pro-vide thrust to lift off the surface of the asteroid and perform a hop to a new location, and to control the robot's attitude during the hop to ensure safe, upright landing. There are 8 nozzles in total, each capable of delivering 1 milli-Newton of thrust. The thruster chip provides control torque on the x and y body axes. The geometry of the nozzles allows for three modes of actuation to control one axis rotation: actuating the inner nozzle, outer nozzle, or both nozzles at the same time.

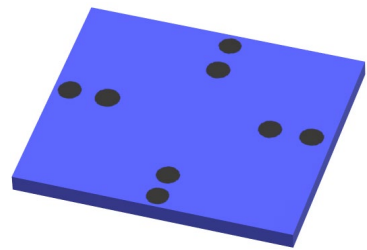

Fig. 5. Thruster Chip

Propellant is stored as a sublimate in a heat-controlled storage chamber (Fig. 6). The sublimate vapor pressure of the propellant is the chamber pressure, analogous to other cold gas and liquid evaporation systems [17-22]. A main valve that provide the main sealing pressure opens to allow for flow to downstream nozzles. Each nozzle is actuated by a simple thruster valve.

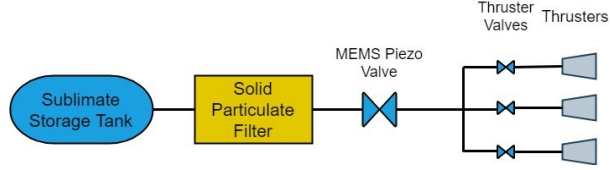

Fig. 6. Propulsion system Block Diagram

As the propulsion system is a chip with thrust only out of plane due to micro-fabrication limits, a reaction wheel is needed to control the z body axis of the robot.

A small reaction wheel is required for z axis control. A commercial off the shelf solution from MAI is taken as a representative solution. The reaction wheel is 8.1 mm in radius, $2.25\times10^{-5}$ kg-m$^2$ inertia, and max spin of 5,000 rpm. This provides sufficient control authority for the robot. By changing the spin rate of the reaction wheel, the change in angular momentum is transferred to the robot to conserve angular momentum.

## 3. Engineering Test Model

The primary purpose of the engineering model is to test path planning and hopping algorithms

### 3.1 Microgravity Simulation

As the gravity environment on a small asteroid is much lower than that experienced on the Earth, a simulation of the microgravity environment is required to accurately assess the robot's dynamics and hopping control.

Three methods were considered for microgravity simulation: tethering, parabolic flights, and helium balloons. Tethering the robot requires significant infrastructure and dynamics analysis to obtain meaningful results. Reduced gravity aircraft in a parabolic flight are both expensive and allow for only short, usually 25 second weightless tests. These short durations are not long enough to complete a full hop, let alone multiple for mapping purposes.

Helium balloons provide a buoyant force to counteract some of the gravitational force experienced by the test model. The mass that can be lifted is equal to the mass of the displaced air by the balloon,

$$m_{buoy} = \rho_{air} V \quad (1)$$

Where $\rho_{air}$ is the density of air and $V$ is the volume of the balloon. In taking account of tShe mass of the helium in the balloon and the mass of the balloon, the mass that can be lifted by the helium balloon is

$$m_{lift} = m_{buoy} - m_{He} - m_{bal} \quad (2)$$

Where $m_{He}$ is the mass of helium in the balloon and

$$m_{He} = \rho_{He} V \quad (3)$$

Where $m_{bal}$ is the mass of the balloon. The buoyancy of the helium balloon should counteract the majority of the weight from the robot. The balloon also augments the inflatable device on AMIGO allowing for a top-mounted stereo camera. From available off the shelf balloons, a 21" mylar spheroid is selected. This is the largest balloon found with a spherical shape with axisymmetric properties.

With a density of air $\rho_{air}$ =1.225 kg/m$^3$ of helium $\rho_{He}$ =0.179 kg/m$^3$, a spherical volume of V = ~0.0795 m$^3$, and mass of the balloon 22 g, the mass that can be lifted by the balloon is $m_{lift}$ = 61.1 g.

Utilizing a drone would also negate gravitational forces should the balloon not provide enough lifting force. A quadcopter would simulate the actuation modes of the bottom mounted thruster chip from Fig. 2, to be discussed in *Section 3.2*.

### 3.2 Avionics and Components

This robot is designed to only use the basic components to simulate hopping and path planning. The included non-structural components are listed in Table 1.

Table 1: Components with their Mass and Cost

| Component | Use | Mass (g) | Cost |
|---|---|---|---|
| Raspberry Pi Zero W | Main Computer | 9.3 | $10.00 |
| eYs3D Stereo Camera | Path Planning | 8 | $179.99 |
| Crazyflie 2.1 | Hopping Actuation | 27 | $195.00 |
| Flow v2 | Motion sensing | 1.6 | $45.00 |
| 1200 mAh LiPo Battery | Power Supply | 23 | $9.95 |
| Power Boost 500 Basic | Volt and Connection Converter | 8 | $9.95 |
| USB 3.1 micro b to A | Camera interfacing | 10 | - |
| A/Micro USB Cable x2 | Power Supply | 10 | $5.90 |
| USB OTG mini-hub | Camera and drone comm | 12 | $4.95 |

The Raspberry Pi Zero W (referred to as Pi from here on out) has been selected as the main computer. This computer runs Raspbian OS with a large library for robotic control, mapping, and communication with 512 MB RAM and a 1 GHz processor. During testing, the Pi will be controlled by SSH from another computer, then left to run on its own during runs to simulate an automated robot system. The Pi receives 5 V power from a micro USB cable, interfaces with the stereo camera from another micro USB ports, and communicates with the drone over UART. UART communication negates the need for a powered USB splitter, as the Pi only has one USB port for sensors. This computer offers sufficient connectivity for all required tasks with user friendly development and a low mass.

Powering the Pi and drone is a 1200 mAh Lithium Ion Polymer (LiPo) battery at 3.7 V. The Pi, however, runs at 5 V and has a power receiving circuit over its USB Power in interface. Thus, the Power Boost 500 Basic is used to convert the battery power to a 5 V USB output. A USB A/Micro cable connects the Pi to this Power Boost.

As the propulsion system requires operation in a vacuum, another hopping mechanism which simulates the discretized actuation points is required. This is accomplished by using a quadcopter with the propellers positioned where the thruster nozzles would be on the bottom of the robot's chassis.

The selected drone is the Crazyflie 2.1 from Bitcraze. This drone was selected as it was the only nano-drone with a programmable API. Most other nano-drones are not directly programmable, so all communication and control must be done with an external computer and controller. Other programmable drones are usually much larger, like from Parrot or DJI, which would negate the usefulness of the helium buoyancy. Communication and power are received over the USB interface.

*3.3 Chassis and CAD Model*

The chassis simulates the volume constraints of a 1U CubeSat with 10x10x10 cm outer dimensions (Fig. 7). The chassis must allow housing for avionics, mechanical interfacing with the Crazyflie, and rigid connection to the helium balloon.

In the goal of minimizing the mass of the system to reduce the load on the drone, balsa wood is chosen as the chassis material. Side panels are 2 mm thick and vertical rods are 1x1x9 cm. The remaining cm in height is allocated for the thickness of the top plate, bottom plate, and screws. The side plates and bottom plate are bonded to the vertical rods with the top open for access to avionics. The top plate is screwed in to threaded inserts bonded in the vertical rods to close the chassis.

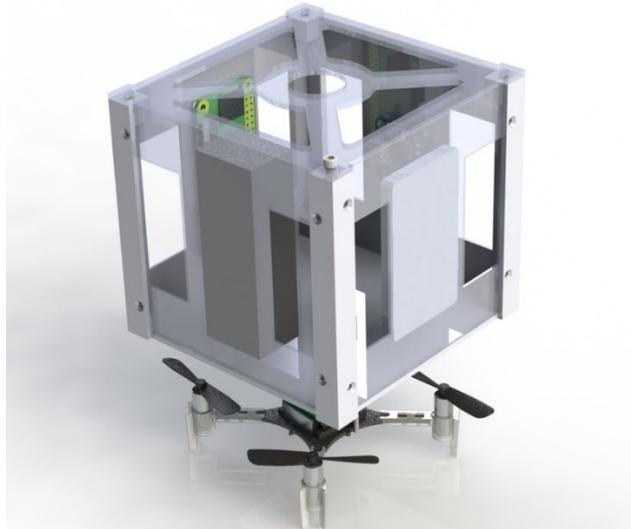

Fig. 7. Chassis and Drone

Upon inflation, the helium balloon is tied to the corners of the chassis (Fig. 8). Strings run from one corner, are tied to the balloon nozzle at the top, and connected to the other corner to keep the balloon against the chassis. An unsecured balloon would cause unknowable disturbances in the robot's operation.

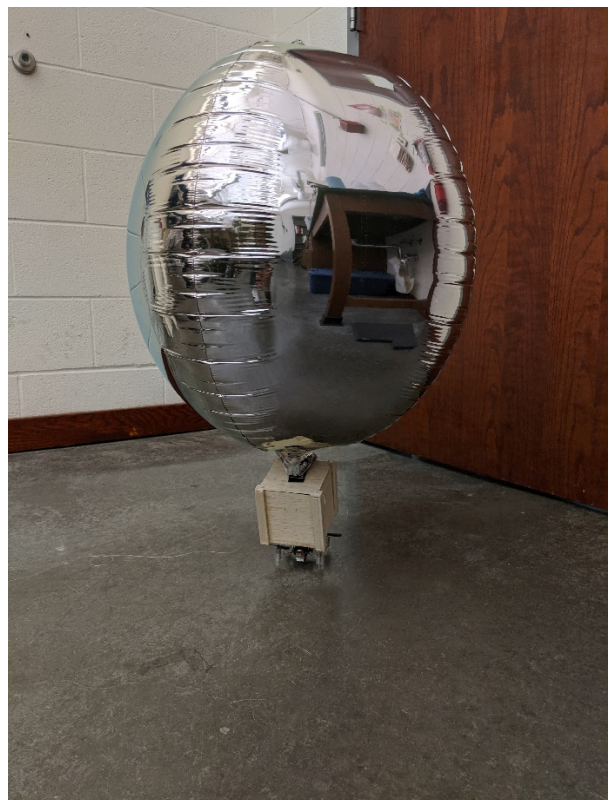

Fig. 8. Test Model with Balloon

## 4. Hopping Program
*4.1 Hop Location Selection*

The output of the stereo camera is a point cloud, color composite image, and depth map with distances in RGB. The on-board image processing from the camera uses the point cloud to create the depth map. An example (Fig. 9) is used with objects placed 1 meter (right box), 2 meters (middle box), and 3 meters (left box) from the robot, with the camera placed on top of the chassis of the robot.

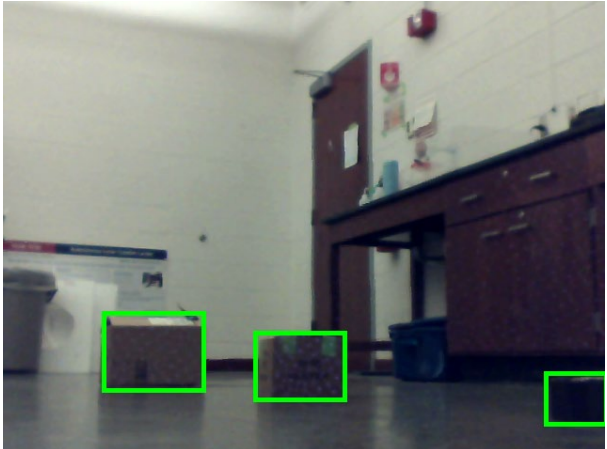

Fig. 9. Color Image of Object Detection

An unprocessed depth map (Fig. 10) is output from the camera software. Far objects are in blue, while the closest objects are in red and white.

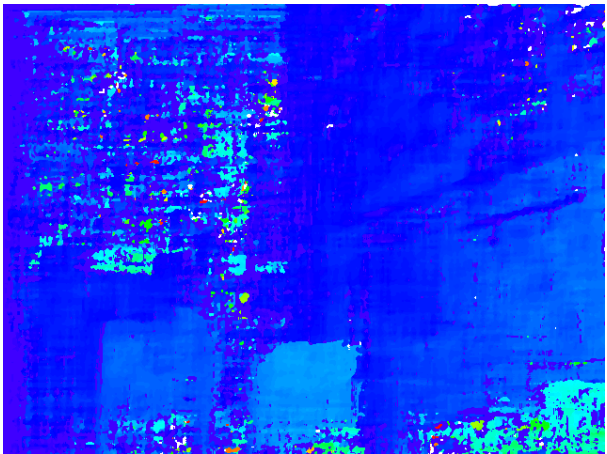

Fig. 10. Unprocessed Depth Map

The images depth map needs to be processed to detect viable paths to hop. This is done by first splitting the depth map into varying regions of distance: "Distant", "Medium", and "Close". In this way, objects at various distances can be classified and noise can be cleaned. From this depth map, it is obvious that adverse and uneven lighting causes imperfect depth information.

A range of RGB values for each distance region is defined. A mask is created to filter out areas not in the distance range in OpenCV, a computer vision library. From the depth map in Fig. 8, each distance field is shown in Fig. 11.

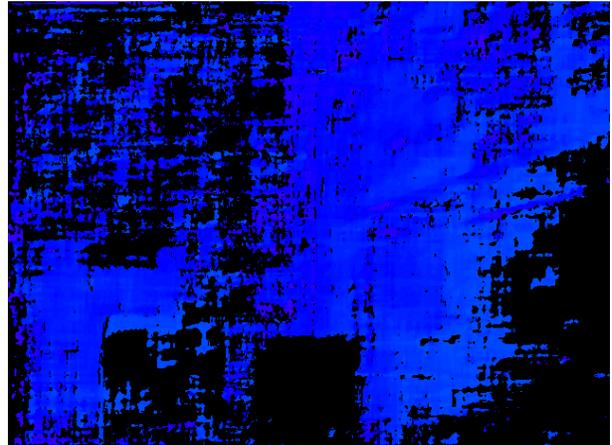

Fig. 11.a. Distant Field Before Processing

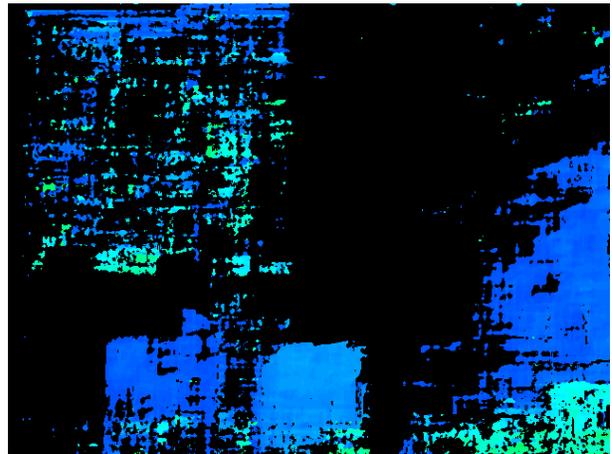

Fig. 11.b. Medium Field Before Processing

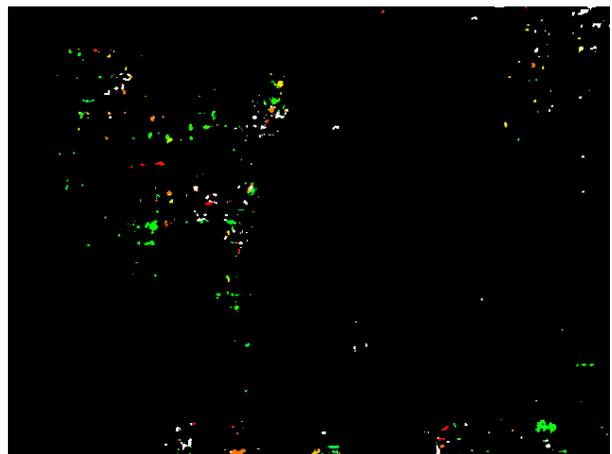

Fig. 11.c. Close Field Before Processing

Contours area created for each object in each distance region. Contours with a very small area are filtered out as

IAC-19-19-B4.8.9x54082                                                                                                                                   Page 6 of 9

noise and places them in the distant field. This tends to clean the distant region that falsely detects small objects. Following this procedure, the distant and medium fields are shown (Fig. 12),

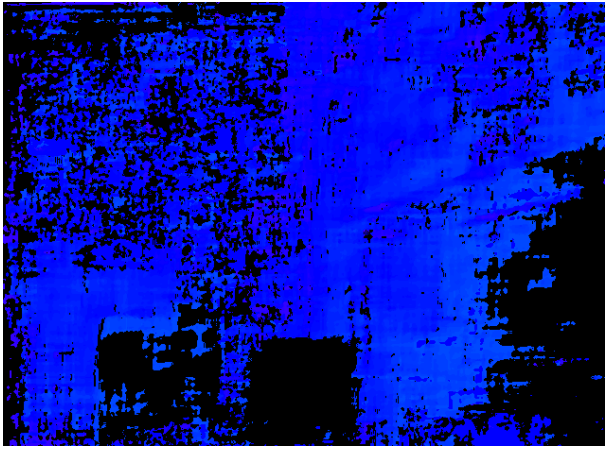

Fig. 12.a. Distant Field After Processing

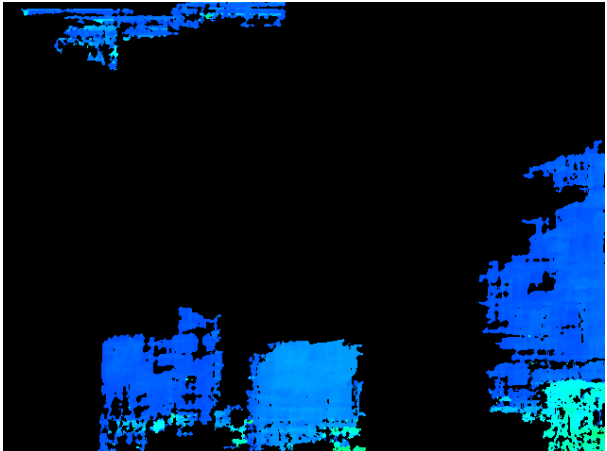

Fig. 12.b. Medium Field After Processing

From the processed images, each of the three obstructions and a cabinet representing a wall have been detected in the "Medium" field. Once the images have been cleaned and obstructions detected, a hop path can be chosen.

Beginning with the distant field, if there is a large enough region with no obstructions, the robot will hop in that direction. The field is split into 80 "slices" vertically, and each slice is evaluated for an obstruction. The slice is an open region if there are no obstructions. Each open region has its neighboring regions checked if they are open. If enough open regions are next to each other, the middle slice is chosen as a valid slice to hop to. If multiple valid slices are found, one is randomly selected. This reduces the likelihood of the robot turning around and hopping in the direction from which it came.

The robot will hop to a maximum distance such that obstructions not previously detected due to their distance will now be in range. Thus, the robot does not over-hop into an unknown region.

If no path is detected, the "Medium" region is analyzed for a path to hop. If a viable direction is chosen, the robot will hop a shorter distance, so it does not encounter a closer obstruction. If no path is selected, the robot will lift straight up, yaw 30° and begin the detection process again. In this way, a new scene is analyzed as the stereo camera is not a true 3D point cloud of the robot's entire surrounding.

### 4.2 Controlling the Crazyflie

The Pi can be used to start the hopping experiment by sending SSH commands over Wi-Fi. This allows for all processes to take place on-board with no external processing required, simulating an autonomous sequence.

To minimize changes to the Crazyflie's firmware architecture, the drone is connected to the Pi via the USB OTG connection for both data and power.

To connect to the drone, the Pi scans for all interfaces and saves the URI of the drone found over USB. As the Flow deck is a motion sensor that integrates x-y movements to obtain the position, errors accumulate that must be reset for each new run. The x-y position is not a true position, but the height is because it uses a range finder.

The Crazyflie is able to receive "set point" commands in a variety of modes: absolute roll-yaw-pitch-thrust, translational velocities and yaw rate, constant height motion, and final position. AMIGO-like quasi-ballistic hops use the velocity set point to simulate an impulsive burn to lift off from the surface. The position is found from the Flow deck to compare accuracy.

These initial velocities are found by a parabolic trajectory to the desired final position. AMIGO uses a single shot algorithm to solve the Lambert Orbital Boundary Value Problem (Eqn. 4) for the irregular, complex gravity fields on small body surfaces.

$$\ddot{r} + 2\omega \times \dot{r} + \omega \times (\omega \times r) + \dot{\omega} \times r = g + d + u \quad (4)$$

Where, $r$ is the position vector, $\dot{r}$ and $\ddot{r}$ are the first and second derivative of the position vector, ω is the angular velocity vector of the asteroid, g is the gravitational acceleration, d is the disturbance acceleration such as SRP and third body perturbations, and u is the control acceleration. For the test model, a simple parabolic trajectory is used due to lack of disturbing forces encountered at asteroids. Once the dynamics of the model are well known and disturbances (mainly drag and non-rigid effects) are characterized, a full simulation using the single shot method can be used.

A different set point is used for the mode of rotating the robot to a new heading. In this mode, the robot hovers 10 cm above the surface and yaws to its new heading.

The hover setpoint is used, with inputs being the desired hover height, yaw rate, and zero velocity in x and y.

## 5. Testing

Testing is done on a simulated asteroid surface with objects placed in the test area to mimic boulders to avoid. Due to the limited test area, the maximum number of hops is set to three.

The position of the robot is logged through the Crazyflie's Kalman estimation algorithm. Height measurements are absolute, while translation is integrated from sensed motion.

## 6. Results

The trajectory of the robot through its three hops is shown in Fig. 13 as a 3D view, while Fig. 14 shows an above view with obstacles approximately where they were placed in the real world.

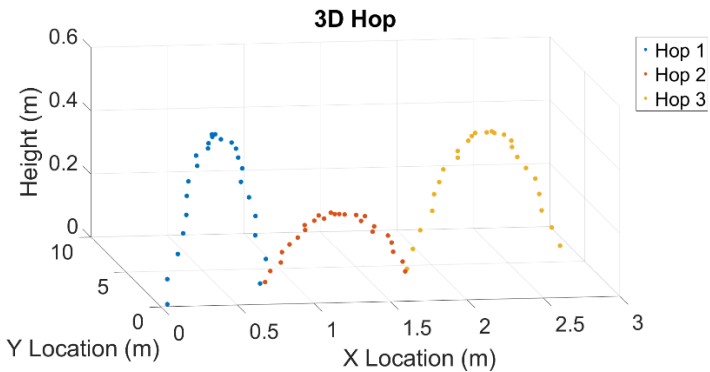

Fig. 13. 3D View of Consecutive Hops

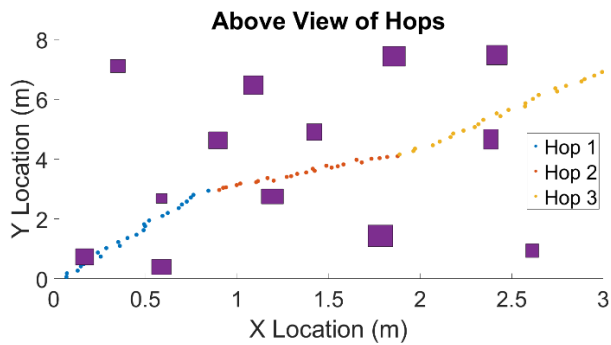

Fig. 14. Aerial View of Hops

The robot is able to successfully navigate between objects on the ground level. By detecting areas to avoid, the robot hops to safe locations and gains a new vantage point from which to observe. This would correlate to AMIGO hopping from destination to destination and collect science data from a more diverse range.

The first and third hops are to the "Distant" field with an unobstructed path to the set maximum distance to hop. The intermediate hop is to the "Medium" field, where there was no area with sufficient area to hop the maximum allowable distance.

Interestingly, the robot's first hop path brought the robot very close to an object. As the robot hopped over the top of it, there was no risk of collision. However, more buffers need to be in place to ensure the robot does not come so close to a forbidden region of the obstacle field.

The shakiness in the plots are from two sources: error accumulation (minor) and wobble during flight (major contributer). The use of the helium balloon causes slight disturbances that causes the robot to wobble. This could cause the robot to deviate from its course and bring it to a non-optimal position in the obstacle field.

## 7. Conclusions

The robot is successfully able to hop using the slicing method of examining a depth map. By first splitting the depth map into three layers, the "Distant", "Medium, and "Close" fields, objects of varying distances can be found and avoided. Each layer is then examined by vertical slices to find open areas.

In searching these open areas as groups, a large enough area to fit the robot through can be found. The algorithm must be updated to ensure the robot does not hop near the edges of objects that disturbances would be sufficient to cause a collision.

Future work will include addition of topography information. Thus far, it has been assumed that there is a flat ground plane that the robot lifts off from and returns to at the same height. On a rough asteroid surface, this is obviously not the case from data gathered by the *Hyabusa-II* and *OSIRIS-Rex* missions.

Another development will be in creating maps of where the robot has explored. Constant feed images and depth maps can be used and stitched together to create a 3D representation of the asteroid surface and obstacles. Currently, snapshots are used when the robot is on the ground to determine the desired hop location, while no images or processing occurs during a hop.

## Acknowledgements

The authors would like to thank Dr. Erik Asphaug and Dr. Stephen Schwartz from the Lunar and Planetary Lab at the University of Arizona for their development of the AMIGO mission concept.

**References**

[1] V. Reddy., et al. "Mineralogy and Surface Composition of Asteroids." Asteroids IV, 2015, doi:10.2458/azu_uapress_9780816532131-ch003.

[2] C. M. Hartzell, D. J. Scheeres, "Dynamics of levitating dust particles near asteroids and the Moon." J. Geophys. Res. Planets, 2013, 118, 116-126, doi:10.1029/2012JE004162.

[3] RM Jones, "The MUSES-CN rover and asteroid exploration mission." In 22nd International Symposium on Space Technology and Science, pages 2403-2410, 2000.

[4] R. Allen, M. Pavone, C. McQuinn, I. A. D. Nesnas, J. C. Castillo-Rogez, Tam-Nquyen, J. A. Hoff-man, "Internally-Actuated Rovers for All-Access Surface Mobility: Theory and Experimentation," IEEE Conference on Robotics and Automation (ICRA), St. Paul, Minnesota, 2012.

[5] B. Hockman, A. Frick, I. A. D. Nesnas, M. Pavone, "Design, Control, and Experimentation of Internally-Actuated Rovers for the Exploration of Low-Gravity Planetary Bodies," Conference on Field and Service Robotics, 2015.

[6] Y. Xu, K. W. Au, G. C. Nandy, H. B. Brown, "Analysis of Actuation and Dynamic Balancing for a Single Wheel Robot" IEEE/RSJ International Conference on Intelligent Robots and Systems, October 1998.

[7] E. Dupius, S. Montminy, P. Allard, "Hopping robot for planetary exploration" 8th iSAIRAS, September 2005.

[8] S. Dubowsky, et al., "A concept Mission: Microbots for Large-Scale Planetary Surface and Subsurface Exploration" Space Technology and Applications International Forum, 2005.

[9] H. Khalita, S. Schwarz, E. Asphaug, J. Thangavelauthum, "Mobility and Science Operations on an Asteroid Using a Hopping Small Spacecrafts on Stilts" 42nd AAS GNC Conference, February, 2019, Breckenridge, CO.

[10] J. Thangavelautham, M. S. Robinson, A. Taits, T. J. McKinney, S. Amidan, A. Polak, "Flying, hopping Pit-Bots for cave and lava tube exploration on the Moon and Mars" 2nd International Work-shop on Instrumentation for Planetary Missions, NASA Goddard, Greenbelt, Maryland, 2014.

[11] H. Kalita, R. T. Nallapu, A. Warren, J. Thangavelautham, "GNC of the SphereX Robot for Extreme Environment Exploration on Mars," Advances in the Astronautical Science, February 2017.

[12] H. Kalita, R. T. Nallapu, A. Warren, J. Thangavelautham, "Guidance, Navigation and Control of Multirobot Systems in Cooperative Cliff Climbing," Advances in the Astronautical Science, February 2017.

[13] Raura, L., Warren, A., Thangavelautham, J., "Spherical Planetary Robot for Rugged Terrain Traversal," Proceedings of the IEEE Aerospace Conference, 2017.

[14] Babuscia, A., Choi, T., Cheung, K., Thangavelautham, J., Ravichandran, M., Chandra, A., "Inflatable antenna for cubesat: Extension of the previously developed s-band design to the X-band,"AIAA Space, 2015.

[15] Wilson, Jim. "What Is NASA's Asteroid Redirect Mission?" NASA, NASA, 16 Apr. 2015.

[16] G. Wilburn, E. Asphaug, J. Thangavelautham, "A Milli-Newton Propulsion System for the Asteroid Mobile Imager and Geologic Observer (AMIGO)", IEEE Aerospace Conference, Big Sky, MT, 2019.

[17] Wu, S.; Mu, Z.; Chen, W.; Rodrigues, P.; Mendes, R.; Alminde, L. TW-1: A CubeSat Constellation for Space Networking Experiments. In Proceedings of the 6th European CubeSat Symposium, Es-tavayer-le-Lac, Switzerland, 16 October 2014.

[18] Bonin, G et al. CanX–4 and CanX–5 Precision Formation Flight: Mission Accomplished! In Proceedings of the 29th Annual AIAA/USA Conference on Small Satellites, Logan, UT, USA, 8–13 August 2015.

[19] Underwood, C.I.; Richardson, G.; Savignol, J. In-orbit results from the SNAP-1 nanosatellite and its future potential. Philos. Trans. R. Soc. Lond. A Math. Phys. Eng. Sci. 2003, 361, 199–203.

[20] Hejmanowski, N.J.C.A.; Woodruff, R.B. CubeSat High Impulse Propulsion System (CHIPS). In Proceedings of the 62nd JANNAF Propulsion Meeting (7th Spacecraft Propulsion), Nashville, TN, USA, 1–5 June 2015

[21] Robin, M.; Brogan, T.; Cardiff, E. An Ammonia Microresistojet (MRJ) for micro Satellites. In Proceedings of the 44th AIAA/ASME/SAE/ASEE Joint Propulsion Conference & Exhibit, Hartford, CT, USA, 21–23 July 2008

[22] Guo, Jian, et al. "In-Orbit Results of Delfi-n3Xt: Lessons Learned and Move Forward." Acta Astronautica, vol. 121, 2016, pp. 39–50., doi:10.1016/j.actaastro.2015.12.003.

[24] Babuscia, A., Sauder, J., Chandra, A., Thangavelautham, J., "Inflatable Antenna for CubeSats: A New Spherical Design for Increased X-band Gain," Proceedings of the IEEE Aerospace Conference, 2017.